\documentclass[prl,twocolumn,superscriptaddress,amsmath,amssymb]{revtex4}
\usepackage{natbib}
\usepackage{graphicx}         
\usepackage{dcolumn}         
\usepackage{bm}                 

\begin{document}

\title{Temperature-dependence of the phase-coherence length in InN nanowires}

\author{Ch. Bl\"omers}
\affiliation{Institute for Bio- and Nanosystems (IBN-1) and JARA
J\"ulich Aachen Research Alliance, Research Centre J\"ulich GmbH,
52425 J\"ulich, Germany}

\author{Th. Sch\"apers}
\email{th.schaepers@fz-juelich.de} \affiliation{Institute for Bio-
and Nanosystems (IBN-1), JARA J\"ulich Aachen Research Alliance,
and Virtual Institute of Spinelectronics (VISel), Research Centre
J\"ulich GmbH, 52425 J\"ulich, Germany}

\author{T. Richter}
\affiliation{Institute for Bio- and Nanosystems (IBN-1) and JARA
J\"ulich Aachen Research Alliance, Research Centre J\"ulich GmbH,
52425 J\"ulich, Germany}

\author{R. Calarco}
\affiliation{Institute for Bio- and Nanosystems (IBN-1) and JARA
J\"ulich Aachen Research Alliance, Research Centre J\"ulich GmbH,
52425 J\"ulich, Germany}

\author{H. L\"uth}
\affiliation{Institute for Bio- and Nanosystems (IBN-1) and JARA
J\"ulich Aachen Research Alliance, Research Centre J\"ulich GmbH,
52425 J\"ulich, Germany}

\author{M. Marso}
\affiliation{Institute for Bio- and Nanosystems (IBN-1) and JARA
J\"ulich Aachen Research Alliance, Research Centre J\"ulich GmbH,
52425 J\"ulich, Germany}

\date{\today}

\hyphenation{InN}

\begin{abstract}
We report on low-temperature magnetotransport measurements on InN
nanowires, grown by plasma-assisted molecular beam epitaxy. The
characteristic fluctuation pattern observed in the conductance was
employed to obtain information on phase-coherent transport. By
analyzing the root-mean-square and the correlation field of the
conductance fluctuations at various temperatures the
phase-coherence length was determined.
\end{abstract}

\maketitle


Semiconductor nanowires are versatile building blocks for the
design of future electronic devices
\cite{Samuelson04,Lu06,Thelander06,Ikejiri07}. This includes, e.g.
nano-scaled transistors \cite{Bryllert06,Li06}, resonant tunneling
devices \cite{Bjoerk02b}, or quantum dot based devices
\cite{DeFranceschi03,Fasth05a}, to name just a few. Among the many
possible materials suitable for semiconductor nanowires, InN is
particularly interesting because of its low energy band gap and
its high surface conductivity
\cite{Liang02,Chang05,Stoica06a,Calarco07}.

At low temperatures electron interference effects often play a
prominent role in the transport characteristics of nanostructures.
Typical phenomena observed in this regime are weak localization,
the Aharonov--Bohm effect, or universal conductance fluctuations
\cite{Beenakker91c,Lin02}. The characteristic length connected to
these effects is the phase-coherence length $l_\phi$, i.e. the
length over which phase-coherent transport is maintained. The
length $l_\phi$ is an important parameter for the design of device
structures based on electron interference.

The analysis of conductance fluctuations is one of the possible
methods, in order to obtain information on $l_\phi$ in
semiconductor nanostructures
\cite{Umbach84,Stone85,Lee85,Altshuler85b,Lee87,Thornton87,Beenakker88a}.
Due to the small dimensions of semiconductor nanowires, often only
a limited number of scattering centers are involved in the
transport. In this case, pronounced fluctuations in the
conductance can be expected, e.g. when the magnet field is varied.
This was indeed observed by Hansen \emph{et al.}\cite{Hansen05}
for InAs nanowires.

In this letter we will exploit the characteristic fluctuation
pattern in the magnetoresistance of InN nanowires, in order to
obtain information on the phase-coherent transport. By analyzing
the root-mean-square and the correlation field of the fluctuation
pattern, the temperature dependence of $l_\phi$ will be
determined.

\begin{figure}[h!]
\includegraphics[width=0.7\columnwidth]{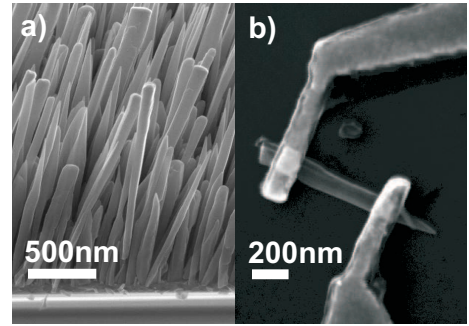}
\caption{Scanning electron micrograph of: (a) the as-grown InN
nanowires used to prepare sample~B, (b) of wire B with the
individually defined Ti/Au contact pads. \label{Fig-photo}}
\end{figure}

Two InN nanowires of different thickness, prepared by
plasma-assisted MBE, were investigated in this study
\cite{Stoica06}. The wires were grown on a Si (111) substrate at a
temperature of 475$^\circ$C under N-rich conditions. For the
growth of the first sample (wire A) a beam equivalent pressure for
In of $3.0 \times 10^{-8}$~mbar was chosen, while for the second
sample (wire B) $7.0 \times 10^{-8}$~mbar was adjusted. As
illustrated in Fig.~\ref{Fig-photo}~a), using this scheme growth
of InN nanowires with a length of approximately 1~$\mu$m was
achieved. From photoluminescence measurements a typical overall
electron concentration of $5 \times 10^{18}$~cm$^{-3}$ was
estimated \cite{Stoica06a}.

For the transport measurements of the InN wires, contact pads and
adjustment markers were defined on a SiO$_2$-covered Si (110)
wafer. The InN nanowires were dispersed on the patterned
substrate, by separating them first from their host substrate in
an acetone solution. Subsequently, a droplet of acetone containing
the detached InN nanowires was put on the patterned substrate
wafer. In the final step, the wires were contacted individually
using Ti/Au electrodes defined by electron beam lithography.
Wire~A had a diameter  of $d=67$~nm with contact pads separated by
$L=410$~nm, while wire B had a diameter of 130~nm with contacts
separated by 530~nm. An scanning electron micrograph of wire B is
shown in Fig.~\ref{Fig-photo}.

The measurements were performed in a He-3 cryostat at temperatures
between 0.6 and 25~K. The cryostat was equipped with a 10~T
magnet. The magnetic field was oriented perpendicular to the axis
of the wires. The magnetoresistance was measured by using a
lock-in technique with an ac bias current of 30~nA.

\begin{figure}[h!]
\includegraphics[width=1.0\columnwidth]{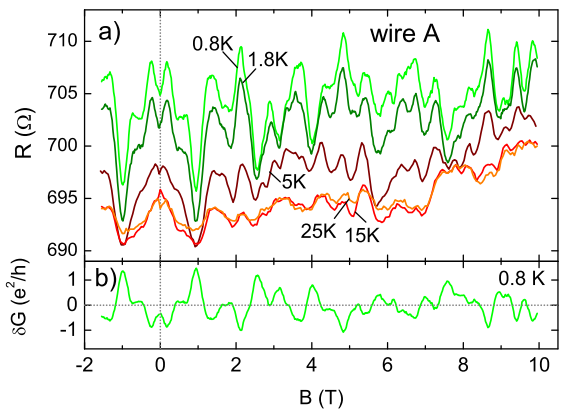}
\caption{(a) Magnetoresistance of wire A at temperatures between
0.8~K and 25~K. The measurements show the total two-terminal
resistance including the contact resistance. (b) Normalized
conductance fluctuations of wire A at 0.8~K.
\label{Fig-wireA-ucf}}
\end{figure}

The fluctuation pattern in the total magnetoresistance of wire A
at various temperatures is shown in Fig.~\ref{Fig-wireA-ucf} a).
The measurements were performed in a magnetic field range from
$-1.5$~T to $10.0$~T. It can be seen clearly that the fluctuation
amplitude decreases substantially if the temperature is increased.
Furthermore, one finds that although the amplitude is damped with
increasing temperature, the pattern itself is reproduced. As can
be seen in Fig.~\ref{Fig-wireA-ucf} a), due to the two-terminal
measurement, the fluctuation pattern is symmetric with respect to
the magnetic field $B$.

In order to analyze the electron interference phenomena in detail,
first, the fluctuations in the magnetoresistance were converted to
the corresponding conductance fluctuations $\delta G$. In
Fig.~\ref{Fig-wireA-ucf} b) the fluctuation pattern of the
normalized magnetoconductance $\delta G$ of wire A at 0.8~K is
shown. Since the measurements were performed in a two-terminal
configuration, first, the contact resistance $R_c$ was subtracted.
The contact resistance was estimated by measuring a number of InN
wires of the same growth run with comparable diameter but
different contact separations. For wire A and B contact
resistances $R_c$ of $(330 \pm 50)\,\Omega$ and $(250 \pm 50)\,
\Omega$, were determined, respectively. After subtracting the
contact resistance, the conductance fluctuations $\delta G=G-G_0$
were extracted by subtracting the slowly varying parabolic
background contribution $G_0$ from the total wire conductance. As
can be seen in Fig.~\ref{Fig-wireA-ucf} b), at 0.8~K  the
conductance fluctuation amplitude is in the order of $e^2/h$.

\begin{figure}[h!]
\includegraphics[width=1.0\columnwidth]{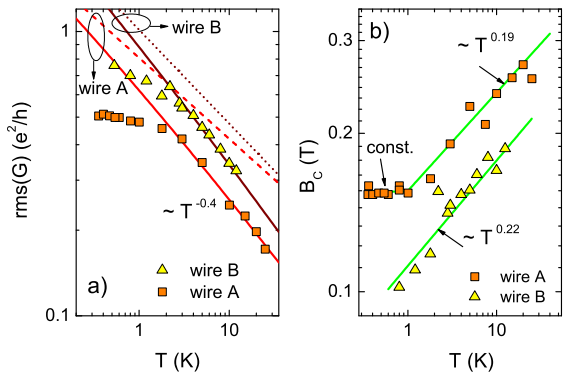}
\caption{(a) root-mean-square (rms) of the conductance
fluctuations of wire A ($\Box$) and B ($\triangle$) as a function
of temperature. The full and dashed lines show the theoretically
expected traces of $\mathrm{rms}(G)$ including and excluding
thermal averaging, respectively. (b) Correlation field $B_c$ vs.
temperature for wire A and B. The full lines represent the
exponential increase of $B_c$. \label{Fig-wireA-rms-Bc}}
\end{figure}

We will now focus on the analysis of the temperature-dependence of
the conductance fluctuations. The magnitude of $\delta G$ can be
quantified by the root-mean-square of the conductance fluctuations
$\mathrm{rms} (G)$, which is defined by $\sqrt{\langle \delta G
^2\rangle}$. Here, $\langle ... \rangle$ represents the average
over the magnetic field. In Fig.~\ref{Fig-wireA-rms-Bc}~a) the
$\mathrm{rms} (G)$ of wire A is plotted as a function of
temperature. Two regimes are revealed: At temperatures below about
1.5~K, $\mathrm{rms}(G)$ tends to saturate, whereas at temperature
above 1.5~K a decrease of $\mathrm{rms}(G)$ proportional
$T^{-0.4}$ is observed. A similar decrease proportional to
$T^{-0.4}$ was found for wire B above $\approx 2$~K, while below
2~K a slightly steeper decrease was observed compared to wire~A.

Before we proceed to discuss the temperature dependence of
$\mathrm{rms}(G)$, we will turn to another important quantity, the
correlation field $B_c$. The correlation field is derived from the
autocorrelation function of the conductance fluctuation which is
defined as $F(\Delta B)=\langle \delta G(B+\Delta B)\delta G
(B)\rangle $.\cite{Lee87} The full width at half maximum of the
autocorrelation function $F(B_c)=\frac{1}{2} F(0)$ defines the
correlation field $B_c$. In Fig.~\ref{Fig-wireA-rms-Bc} b) the
temperature dependence of $B_c$ is plotted for both wires. For
wire~A the correlation field remains constant for temperatures
smaller than 1~K, while for larger temperatures $B_c$ increases
proportional to $T^{0.19}$. In contrast, for wire B the
correlation field $B_c$ monotonically increases with $T^{0.22}$ in
the whole temperature range.

\begin{figure}[h!]
\includegraphics[width=1.0\columnwidth]{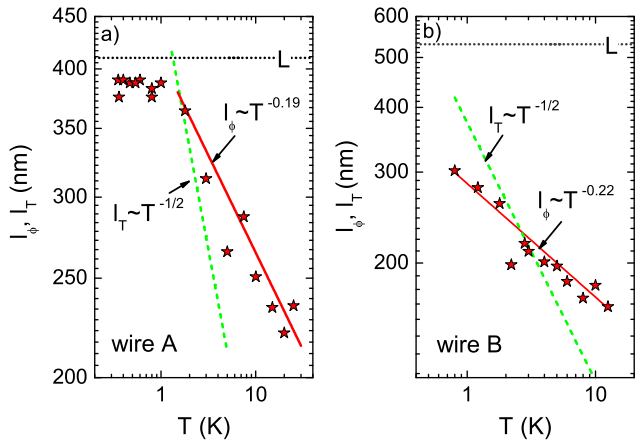}
\caption{(a) Phase-coherence length $l_\phi$ of wire~A determined
from $B_c$ as a function of temperature ($\star$). The solid line
represents the $T^{-0.19}$-dependence of $l_\phi$ above 1.5~K. The
dashed line corresponds to the expected temperature dependence of
$l_T$. The wire length $L$ is indicated by the dotted line. (b)
Corresponding values for wire B. \label{Fig-compare-lphi}}
\end{figure}

In order to estimate the temperature dependence of $l_\phi$, we
focus the analysis of $B_c$. No attempt was made to determine
$l_\phi$ directly from $\mathrm{rms}(G)$, because
$\mathrm{rms}(G)$ depends on the interplay between two parameters:
$l_\phi$ and $l_T$ \cite{Lee87}. The thermal diffusion length
$l_T=\sqrt{\hbar \mathcal{D}/k_BT}$, with $\mathcal{D}$ the
diffusion constant, is a measure for the thermal broadening. In
contrast, $B_c$ does not depend on thermal broadening effects,
i.e. on $l_T$ \cite{Beenakker88a}. Since in our wires the diameter
$d$ exceeds the elastic mean free path $l_e$, $l_\phi$ was
determined from $B_c$ using the expression for the diffusive
regime: $l_\phi \approx \Phi_0/B_cd$, with $\Phi_0=h/e$ the
magnetic flux quantum \cite{Lee87,Beenakker88a}. The results for
sample A are shown in Fig.~\ref{Fig-compare-lphi}~a).
For $T<1.5$~K the calculated values of $l_\phi$ saturate at about
the wire length $L$, indicating that phase coherence is maintained
over the complete length. Above 1.5~K, $l_\phi$ decreases with
increasing temperature following a dependence proportional to
$T^{-0.19}$. Using the same procedure as described above, $l_\phi$
was determined from $B_c$ for wire~B, as well. As can be seen in
Fig~\ref{Fig-compare-lphi}~b), $l_\phi$ monotonously decreases
with temperature following a dependence proportional to
$T^{-0.22}$ in the whole temperature range. Consequently, one can
state that for this wire $l_\phi$ is always smaller than $L$ in
the temperature range considered here. For both wires the
temperature dependence of $l_\phi$ is slightly smaller than the
theoretically expected dependence proportional to $T^{-1/3}$
\cite{Altshuler82}.

Using the interpolation formula derived by Beenakker and van
Houten \cite{Beenakker88a}:
\begin{equation}
\mathrm{rms}(G)=\alpha \frac{e^2}{h}
\left(\frac{l_\phi}{L}\right)^{3/2}\left[1+\frac{9}{2\pi}\left(
\frac{l_\phi}{l_T}\right)\right]^{-1/2} \; , \label{Eq-1}
\end{equation}
the temperature dependence of $\mathrm{rms}(G)$ was estimated. The
formula is valid for $l_\phi \approx l_T <L$. Ideally, the
constant $\alpha$ has a value of $\sqrt{6}$. The calculated values
for wire A and B using Eq.~(\ref{Eq-1}) are shown in
Fig.~\ref{Fig-wireA-rms-Bc}. One, finds that at $T>1$~K the
experimental values of $\mathrm{rms}(G)$ vs. $T$ fit very well to
the theoretical curves. The values of $\alpha$ with 0.38 and 1.32
for wire A and B, respectively, deviate to some extent from the
theoretically expected value. This can be attributed to
uncertainties in the determination of $R_c$ affecting the
amplitude of $\mathrm{rms}(G)$. For comparison, the
$\mathrm{rms}(G)\propto (l_\phi/L)^{3/2}$ curves, representing the
case of neglected thermal averaging, are also shown in
Fig.~\ref{Fig-wireA-rms-Bc}. One finds that in this case the
decrease of $\mathrm{rms}(G)$ is too small. The fact that thermal
smearing has to be considered is also supported by comparing $l_T$
to $l_\phi$ (cf. Fig.~\ref{Fig-compare-lphi}). Here, one finds
that $l_\phi$ and $l_T$ are in the same order of magnitude. At
lower temperatures ($T<1$~K) the smaller slope of
$\mathrm{rms}(G)$ can be explained by the reduced effect on $l_T$
and to the fact that $l_\phi$ exceeds or approaches $L$.

In summary, the phase coherence length $l_\phi$ of InN nanowires
was determined by analyzing the characteristic conductance
fluctuation pattern. For the shorter wire it is found that at low
temperatures $l_\phi$ exceeds the wire length $L$, while at
temperatures above 1.5~K $l_\phi < L$ and continuously decreases
with increasing $T$. In contrast for the longer wire (sample B)
$l_\phi$ was smaller than $L$ in the complete temperature range.
Our investigations demonstrate, for short InN nanowires phase
coherent transport can be maintained along the complete length.

\end{document}